\documentclass[aps,prl,twocolumn]{revtex4}
\usepackage{amssymb}
\usepackage{amsfonts}
\usepackage{amsmath}
\usepackage{graphicx}

\begin{document}
\title{Topological Surface Superconductivity Induced by Hydrostatic Pressure-Enhanced Antisymmetric Spin-Orbit Coupling in Non-Centrosymmetric Superconductor PbTaSe$_2$}

\author{Cong Ren$^{1,2\ast}$, Hai Zi$^{1,2}$, Yu-jia Long$^{2}$, Lin-xiao Zhao$^{2}$, Xing-yuan Hou$^2$, Huan-xing Yang$^{2}$, Yi-feng Yang$^{2,3}$, Lei Shan$^{2,3}$, Zhi-an Ren$^{2,3}$, Jian-qi Li$^{2,3}$, Jiang-ping Hu$^{2,3}$, Peng Xiong$^{4}$, Geng-fu Chen$^{2,3\dag}$}

\address{$^1$ Physics Department, School of Physics and Astronomy, Yunnan University, Kunming 650500, China}

\address{$^2$ Beijing National Laboratory for Condensed Matter Physics, Institute of Physics, Chinese Academy of Science, Beijing 100190, China}

\address{$^3$ School of Physical Sciences, University of Chinese Academy of Sciences, Beijing 100190, China}

\address{$^4$ Physics Department, Florida State University, Tallahassee, FL 32306 USA}

\begin{abstract}
A notable characteristic of PbTaSe$_2$, a prototypical noncentrosymmetric (NCS) superconductor, is that its superconductivity can be modulated through a structural transition under hydrostatic pressure [Phys. Rev. B 95, 224508 (2017)].  Here we report on simultaneous pressure-sensitive point-contact Andreev reflection (PCAR) spectroscopy and bulk resistance measurements on PbTaSe$_2$, to elucidate the nature of the surface and bulk superconductivity and their evolution with hydrostatic pressure.  It is found that in high pressure region the superconducting gap opening temperature $T_c^A$ is significantly lower that the bulk resistive transition temperature $T_c^R$, revealing a clear experimental signature of surface-bulk separation associated with enhanced antisymmetric spin-orbit coupling (ASOC).  The PCAR spectra, reflecting the superconducting surface state, are analyzed with the Blonder-Tinkham-Klapwijk theory, yielding an isotropic $s$-wave full BCS-gap in the strong coupling regime.  Analysis based on a modified McMillan formula indicates a sizable coupling strength contributed from ASOC for the superconducting surface state.  These results suggest the coexistence of full gap $s$-wave superconductivity and topological surface states in PbTaSe$_2$, indicating that this NSC with significantly enhanced ASOC may offer a solid platform to investigate the topological aspect in the superconducting condensate.
\end{abstract}

\maketitle

The discovery of superconductivity in topological matters has stimulated intense interest in topological superconductors (TSCs) that may harbors exotic electronic excitation, such as Majorana Fermion zero modes.  Analogous to topological insulators where strong spin-orbit coupling (SOC) causes bulk band inversion, creating a surface state with a Dirac band dispersion, TSCs are characterized by nontrivial symmetry-protected surface states hosting a Dirac dispersion as well as helical spin polarization \cite{a1,a2,a3,a4,a5,a6,a7,a8,a9,a10}.  For realization of such intriguing superconducting topological surface state (TSS), intrinsic TSCs exhibiting chiral $p$-wave pairing are a natural choice \cite{b11,b12,b13,b14}. However, $p$-wave superconductivity is extremely rare and fragile in real materials, and the same is expected for the chiral $p$-wave pairing-engendered surface states.  Two alternately pathways to more robust superconducting TSS were proposed: (i) Superconducting surface state induced by proximity coupling between a strong topological insulator and an $s$-wave superconductor \cite{a2}. The Bi$_2$Se$_3$/NbSe$_2$ heterostructures are a notable example \cite{15JiaJF,16,17LiQ}; (ii) Intrinsic proximity-induced surface states in a superconductor hosting strong SOC.  For the latter case, a noncentrosymmetric superconductor (NCS) under certain circumstances provides a promising route to the physical realization of a TSS \cite{a6,a10,18NSC} In NCS, the absence of a center of inversion symmetry in the crystal structures introduces an asymmetric potential gradient and, consequently, an antisymmetric spin-orbit coupling (ASOC) \cite{18,19}. The ASOC can lift the twofold spin degeneracy of the electron band, giving rise to TSS and allowing an admixture of spin-singlet (even parity) and spin-triplet (odd parity) pairing components, with a mixing ratio tunable via the ASOC strength.  Hence, an NCS with predominant spin-triplet pairing can be a bulk TSC, whereas an NCS with spin-singlet pairing in the bulk can still lead to a TSC on the surface, if fully gapped superconductivity is induced on the TSS.  An NCS with a superconducting TSS is therefore an intriguing platform for TSC.  Partly motivated by these theoretical proposals, a number of NCSs have been synthesized recently to provid many potential platforms to investigate the ASOC-associated topological superconductivity \cite{20LiPdPtB,21LaNiC,Cava-TRB,22BiPd,23PbTaSe}.

Among NCSs, chemically stoichiometric PbTaSe$_2$ has emerged as a strong candidate for TSC \cite{23PbTaSe}.  Surface-sensitive angle-resolved photoemission spectroscopy (ARPES) revealed that PbTaSe$_2$ possesses a fully spin-polarized Dirac surface state \cite{24ARPES}, consistent with the theoretical calculations \cite{23PbTaSe}. A recent scanning tunneling microscopy/spectroscopy study \cite{25STM} reported two spin-polarized Dirac TSSs on the Pb-terminated surface that also be fully gapped at low temperature together with the bulk counterpart. The surface states exhibit essentially identical superconductivity, i.e., the same gap size and critical temperature $T_c$ as those of the bulk state. This is in contrast to the theoretical expectation that the superconducting gap or/and $T_c$ on the TSS should be different from that of the bulk \cite{26theory-Jap}. In light of the absence of a distinct interface between the bulk and surface states in PbTaSe$_2$, it remains an experimental challenge to discern their contributions to superconductivity in tunneling spectroscopies.  As far as superconducting order parameter is concerned, there exists no clear experimental evidence for a superconducting surface state which is unambiguously distinct from the bulk state. One possible reason for the absence of a distinct superconducting surface state is that the extent of inversion symmetry breaking in PbTaSe$_2$ under ambient conditions may not be significant enough to induce a large SOC and a discernible surface-bulk separation.

\begin{figure}
\includegraphics[scale=0.3]{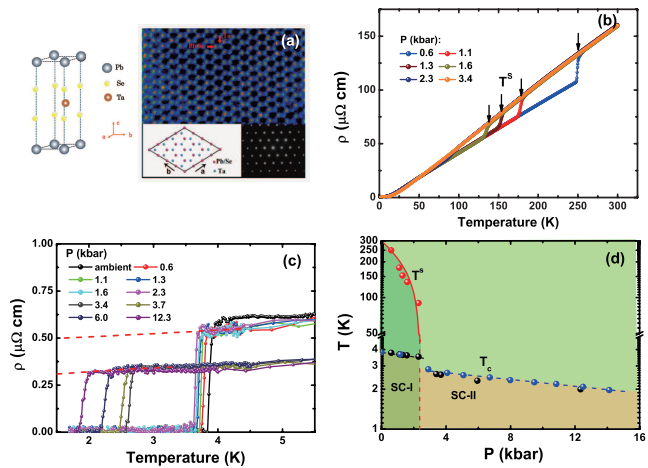}
\caption{(Color online) (a) Characterization of the crystal structure of the PbTaSe$_2$ samples: The red arrows in the main panel point to Pb(Se) and Ta atoms locations in the high-resolution scanning transmission electron microscopy (STEM) image. The scale bar is 2 nm. (b) In-plane $\rho $ verse $T$ in the full measurement temperature range under various hydrostatic pressures. (c) In-plane $\rho $ vs. $T$ under various $P$s in the superconducting transition region. The red dashed lines are the extrapolation to $T$=0; (d) The resulting $P$-$T$ phase diagram for PbTaSe$_2$ on a semi-log scale.  The dashed lines are guides to the eye.}
\end{figure}

Recently, pressure-dependent electrical transport measurements in PbTaSe$_2$ revealed two distinct types of superconductivity at high hydrostatic pressures associated with a subtle structural transition \cite{29Canfield,30XuXF}. The observations point to an intriguing opportunity for exploring ASOC-enhanced and bulk-separated surface topological superconductivity in this high-pressure phase. In this Letter, we performed simultaneous planar point-contact Andreev reflection (PCAR) spectroscopy and bulk resistance measurements on PbTaSe2 single crystals under hydrostatic pressures. In the high-pressure phase, the PCR spectra reveal a superconducting gap opening temperature substantially lower than the bulk resistive transition temperature. The results present compelling evidence for distinct surface-bulk separated superconducting states, likely originated from the enhanced ASOC strength due to the hydrostatic pressure.

High quality single crystals of PbTaSe$_2$ were grown by chemical vapor transport by using iodine as a transport agent.  The details of single crystal growth and physical property characterization were reported elsewhere \cite{29Canfield,31Taiwan,32LongYJ}. The crystal structure and its noncentrosymmetric nature have been examined by scanning transmission electron microscopy (STEM) along the [001] zone-axis direction with high resolution.  In Figure 1(a), the large brighter dots in the STEM image correspond to the projections of the Pb/Se columns, and the small bright dots are the atomic images of the Ta columns. The image and corresponding diffraction pattern clearly evidence the high quality of the sample and its non-centrosymmetric structure.

The temperature dependence of the resistivity $\rho (T)$ of PbTaSe$_2$ single crystal under various pressures ($P$s) is shown in Figure 1(b).  These $P$-dependent $\rho(T)$ curves are typical of those reported in the literature \cite{29Canfield,30XuXF}: Upon the application of pressure, a steep drop of $\rho$ occurs at temperature $T^s$ indicated by the arrows, clearly dividing $\rho(T)$ curves into two branches of low-$\rho$ and high-$\rho$ states.  With increasing $P$, $T^s$ decreases and the $\rho$ drop is gradually suppressed and disappears at $P>2.3$ kbar.  A detailed experiment by Kaluarachchi \textit{et al.}, reveals a thermal hysteresis between high-$\rho$ and low-$\rho$ phases at around $T^s$ for $P<2.4$ kbar, implying a subtle crystal structure modification from the high-$\rho$ $P6m2(1e)$ phase to the low-$\rho$ $P\bar{6}m2$ phase \cite{29Canfield}. In comparison to the $P\bar{6}m2$ phase, the high-$\rho$ $P\bar{6}m2(1e)$ phase displays a sizable contraction in the $c$-axis while the $a$-axis is slightly expanded, leading to an enhanced non-centrosymmetry.

This normal-state $\rho$ transition at $T^s$ in PbTaSe$_2$ shows a distinct correlation with its superconductivity at low-$T$.  As shown in Figure 1(c), a blow-up of resistance in superconducting transition regime clearly splits into two superconducting-transition branches.  In the low-$P$ region (high-$\rho$ branch, $P<2.3$ kbar), the sample shows sharp superconducting transitions with $T_c$ varying little with pressure (SC-I).  In contrast, pressures above 2.9 kbar result in a substantial decrease in the residual resistance, accompanied by large suppression of $T_c$, suggesting a different superconducting state under high pressure (SC-II).  These results of $T^s$ and $T_c$ are used to construct a $T-P$ phase diagram shown in Fig. 1(d), which indicate two distinct superconducting phases separated by a structural transition. First-principle calculations also point to the structural origin for this Lifshitz transition \cite{30XuXF} without any change of global symmetry.  Another piece of evidence for enhanced ASOC in SC-II comes from the observation of much lowered $T_c$  in this high $P$ phase (SC-II) with otherwise similar parameters, suggesting that a strong ASOC is detrimental to superconductivity \cite{33Lonzarich,34ZhengGQ}.

\begin{figure}
\center\includegraphics[scale=0.3]{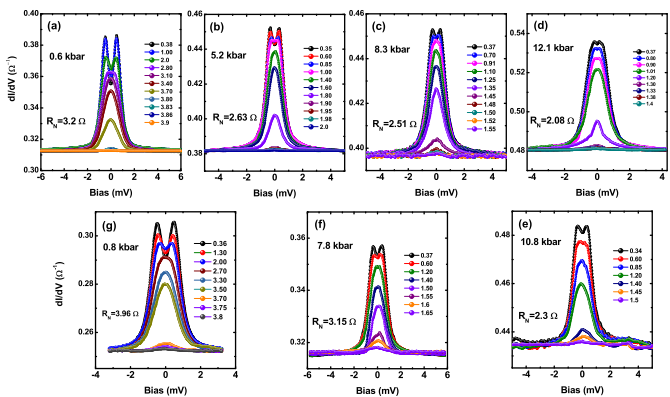}
\caption{(Color online) Evolution of the planar point-contact Andreev conductance spectrum $G$=d$I$/d$V$ vs. bias voltage $V$ as a function of $T$ for various hydrostatic pressures. Shown in panels (a)-(d) for increasing pressure, and in panels (e)-(g) are for decreasing pressures. The labeled junction resistance $R_N$ is defined based on the conductance values at high bias voltages in the normal state. }
\end{figure}

Significantly, the simultaneous point contact Andreev reflection (PCAR) spectroscopy under pressure on the same sample provides unique new insights into the effect of the enhanced ASOC on the nature of the NCS superconductivity across the critical pressure of $P_c \simeq 2.5$ kbar. The high-pressure PCAR measurement on PbTaSe$_2$ crystals were made possible by ``soft'' point contacts made using a thick silver paste bonding with Pt wires to the flat and shiny surface cleaved along the $c$-axis of PbTaSe$_2$  crystals.  Details of the point contacts fabrication can be found in Supplemental Materials SI. It was demonstrated that the contacts thus made ensure a much higher mechanical and thermal stability, and more importantly, these planar contacts have the advantage of avoiding inhomogeneous local pressure effects induced by a metal-tip \cite{35IndianBiPd,36WangJCr3As2}.

Andreev reflection (AR) is the process at a normal metal/superconductor interface that converts the quasiparticle current in normal metal into supercurrent in the superconductor. Experimentally, the Andreev spectroscopy can detect the surface-state superconductivity due to its surface-sensitive nature. Figure 2(a)-(d) show the raw conductance curves, $G(V)=dI(V)/dV$, of a $c$-axis PbTaSe$_2$/Ag point contact as a function of $T$ under several hydrostatic pressures up to 12.1 kbar. To ensure the reliability and reproducibility of the $P$-dependent spectra, we measured the conductance spectra with unloading $P$. As shown in Fig. 2(e)-(g), the $G(V)$ curves show reproducible spectroscopic features, supporting the validity of the $P$-dependent PCAR spectra. Several features of the conductance are worth noting: (i) The spectra show a systematic evolution with the variation of $P$ across the $P_c$. The pronounced double-peak at the gap edges at low $P$ is gradually smeared out with increasing $P$, implying increased junction transparency and dominance of the AR process. (ii) $G(V)$ curve becomes flat and featureless as $T$ approaches $T_c$, and overlaps at high bias voltage with the normal-state curves. The well-defined double peaks at the gap-edges at the lowest $T$ is a tell-tale signature of an $s$-wave-like order parameter in the superconductor and contributions from both Andreev reflection and quasiparticle tunneling (normal reflection). At $P > P_c$, $G(V)$ spectrum shows a zero-bias conductance dip, instead of a peak. A zero-bias conductance peak in the junction spectrum is widely regarded as a manifestation of a $p$-wave nodal-gap in the superconducting order parameter. The persistent presence of the zero-bias conductance dip in the high-pressure region is strong evidence against the dominance of a $p$-wave order parameter in the SC-II regime.

\begin{figure}
\center\includegraphics[scale=0.4]{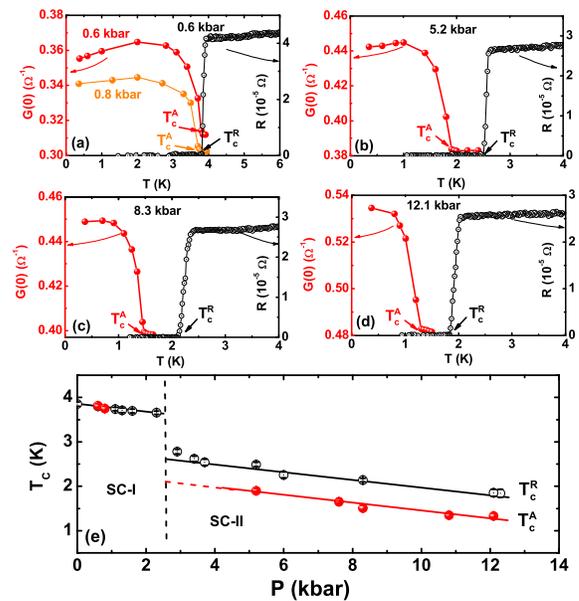}
\caption{(Color online) (a)-(d) Red solid dots: $T$-dependence of the zero-bias conductance, $G(0)$, of the PbTaSe$_2$/Ag planar point-contact spectra under different pressures. Black open circles: The $R-T$ curves simultaneously measured on a PbTaSe$_2$ crystal from the same patch. (e) The superconducting ordering temperature $T_c$ of PbTaSe$_2$ crystals in a $T-P$ phase diagram. The red solid circles are $T^A_c$, and the black open circles are $T^R_c$. The black and red solid and dotted lines are guide to the eye for $T^R_c$ and $T^R_c$ respectively.}
\end{figure}

The simultaneous determination of the $P$-dependent point-contact conductance spectra and the bulk transport properties offer the intriguing possibility of discerning surface-bulk separation by examining the superconducting order temperatures of the surface and bulk states. In principle, the Andreev reflection is manifested as the conductance enhancement in $G(V)$, especially the zero-bias value $G(0)$, above the normal-state counterparts. A superconducting gap opening/closing temperature, $T^A_c$, is defined as the temperature at which the Andreev conductance features disappear and the conductance spectrum becomes indistinguishable from that of the normal state. In this criterion, $T^A_c$ falls within the width of the superconducting fluctuation region in which defines the bulk superconducting transition temperature $T^R_c$ in $\rho(T)$ curve for conventional superconductors.  In Figure 3, we present both the zero-bias PCAR conductance $G(0)$ and the simultaneously measured bulk resistance of the PbTaSe$_2$ crystal as functions of temperature under several $P$s. For all $P$s, $G(0)$ shows a clear transition from sharp variation with $T$ to almost $T$-independent, making the identification of $T^A_c$ straightforward. As shown in Fig. 3(a), in the low-$P$ regime ($P\simeq 0.6$ and 0.8 kbar), the gap opening temperature ($T^A_c \simeq 3.8$ K) identified from $G(0)$ exactly matches the resistive transition temperature $T^R_c$.  In contrast, above $P_c$ in the SC-II state, as shown in Figs. 3(b), (c), (d) at $P$=5.2, 8.3, 12.1 kbar, respectively, there exist a clear bifurcation of $T^A_c$ from $G(0)$ and $T^R_c$ from $R(T)$. For example, for $P$=5.2 kbar [Fig. 3(b)], $T^A_c \simeq 1.90$ K, far below the $T^R_c=2.50$ K of the $R(T)$ curve. In Fig. 3(e), we construct a superconducting $T_c-P$ phase diagram based on the data of $T^A_c$ and $T^R_c$ from these measurements, and those of $T^R_c$ obtained from the $R(T)$ curves in Fig. 1(d). Here, we emphasize that the simultaneous PCAR and $R-T$ measurements in our experiment eliminated random errors in the pressure determination. It is evident that in the SC-II region characterized by an enhanced ASOC, the $T_c-P$ curve is split into two distinct branches, i.e., $T^A_c-P$ and $T^R_c-P$ lines, in contrast to the case in the SC-I state ($P < P_c$) with weaker ASOC strength. This distinction of $T_c$ determined from bulk resistivity and PCAR spectrum measurements is a strong indicator of a superconducting surface state proximity-induced by the enhanced ASOC strength. Remarkably, this is regarded as a key experimental signature of surface-bulk separation in momentum space. Considering the effects of spin-band splitting due to strong ASOC, theoretical calculations by Schnyder \textit{et al.}, revealed that for NCSs there exists a finite region of gapless nodal superconducting phase between the fully gapped topologically trivial (bulk state) and nontrivial (surface state) phases \cite{39Timm,40Eschrig}.

\begin{figure}
\center\includegraphics[scale=0.5]{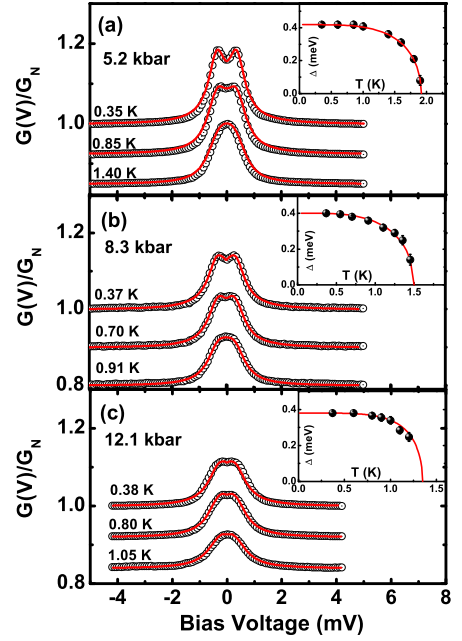}
\caption{(Color online) Normalized PbTaSe$_2$/Ag planar point contact Andreev conductance spectra, $G(V)/G_N$, under three $P$s as functions of $T$. The red solid lines are BTK-model fits based on a single $s$-wave gap. The upper $G(V)/G_N$ curves with their fits are offset for clarity. Insets: The superconducting gap values $\Delta$ from the BTK fits as functions of $T$. The red solid lines are BCS fits.}
\end{figure}

To quantitatively assess the nature of the superconductivity in the TSS of PbTaSe$_2$ in the SC-II state, we determine the gap function $\Delta$ by fitting the PCAR spectra to a generalized Blonder-Tinkham-Klapwijk (BTK) model \cite{41BTK}.  Examples of the normalized $G(V)$ curves and their BTK fits at the selected $T$s are shown in the main panels of Figs. 3(a), (b) and (c) for $P$=5.2, 8.3, 12.1 kbar, respectively. The use of single isotropic $s$-wave gap in the BTK model (red lines) yields best fits to the data, which capture very well all the important features of the experimental $G(V)$ curves. The analyses yield a set of $T$-dependent fitting parameters, especially the gap size, $\Delta(T)$, at different temperatures.  The insets of Fig. 3(a), (b), and (c) are the extracted gap values as functions of $T$ for the corresponding $P$s. The resulting $\Delta(T)$ are well-described by the BCS-gap function: $\Delta(T)=\Delta_0\tanh(\alpha \sqrt{T_c/T-1})$ where $\alpha$ is the coupling strength, shown as the red solid lines in the insets.  Using this formula, we obtained the zero-$T$ gaps $\Delta_0$=0.43, 0.40, 0.37 meV, and $T_c$=1.88, 1.50, 1.35 K, for $P$=5.2, 8.3 and 12.1 kbar, respectively. It is worth noting that in this formula, the fitting parameter $T_c$ is the critical temperature for gap opening/closing ($\Delta$=0). These $T_c$ values are fully consistent with $T^A_c$ determined from the zero-bias conductance $G(0)$ (Fig. 3).

In addition, the ratio $2\Delta_0/k_B T_c$ is a measure of the coupling strength of the superconducting state. For a BCS superconductor in the weak coupling limit, $2\Delta_0/k_B T_c \simeq 3.52$. Here, for PbTaSe$_2$ with the enhanced ASOC in the SC-II region, the ratio increases with $P$, reaching 6.2 at 12.1 kbar.  This is in contrast to the case of a conventional strong-coupling superconductor, for example, in Pb $2\Delta_0/k_B T_c$=4.66 and the value decreases gradually with increasing $P$ \cite{42RenAPL}. Our experimental results strongly suggest that the proximity-induced superconductivity in the TSSs of PbTaSe$_2$ is in the strong coupling regime, in accord with a theoretical prediction that a larger superconducting gap in the TSS may be induced by surface Dirac fermions \cite{26theory-Jap,theory-Sarma}.

Finally, to account for the differences of critical temperatures ($\delta T_c\simeq 0.6$ K) between the bulk and the surface superconductivity in PbTaSe$_2$, we assume that the two superconducting states have the same lattice vibration frequency, i.e., Debye temperature $\theta_D$ under the same $P$. Using a modified McMillan expression which takes into account of SOC effects \cite{43spinMc,44spinMc,45spinMc}
\begin{equation}
T_c=\frac{\theta_D}{1.45}\exp[-\frac{1.04(1+\lambda_{eff})}{\lambda_{eff}-\mu^{\ast}(1+0.62\lambda_{eff})}],
\end{equation}
where
\begin{equation}
\lambda_{eff}=\frac{\lambda_{e-ph}}{1+\lambda_{SOC}}.
\end{equation}
Here $\lambda_{SOC}$ is a contribution arising from ASOC, $\lambda_{e-ph}$ the electron-phonon coupling constant, and $\mu^{\ast}\simeq 0.15$ eV the renormalized Coulomb parameter from which does not vary significantly for the surface-state superconductivity \cite{46Duan}.  Using $\theta_D\simeq 217$ K, $T_c^R \simeq 2.50$ K and $T_c^A \simeq 1.90$ K at $P\simeq 5.2$ kba \cite{29Canfield}, it is obtained that $\lambda_{e-ph}=0.62$ and $\lambda_{ASOC}\simeq 0.07$.  Further theoretical calculations for topological superconductivity which is mediated by electron-phonon interactions should take into account of such sizable spin-orbit coupling strength, such as the case of a few layers of PbTaSe$_2$ in which both gap amplitude and $T_c$ should be the functions of the thickness of PbTaSe$_2$ layers \cite{46Duan}.

In summary, we have performed simultaneous pressure-dependent point-contact spectroscopy and bulk resistance-temperature measurements under hydrostatic pressures to investigate the effect of ASOC on the superconductivity of PbTaSe$_2$. With increasing pressure, PbTaSe$_2$ experiences a structural Lifshitz transition with inversion symmetry breaking, leading to an enhanced ASOC in the high-$P$ phase. In the high-$P$ region, the superconducting-gap opening temperature $T^A_c$ of the surface state is found to be consistently lower than the bulk resistive transition temperature $T^R_c$. The observations are interpreted as evidence for proximity-induced superconducting TSSs in the ASOC-split band.  Applying the modified McMillan equation to the TSS superconductivity reveals a sizable coupling strength originated from the enhanced ASOC. The definitive observation of the surface-bulk separated superconductivity in PbTaSe$_2$ indicates that it is an intrinsic TSC under high pressure. Further experiment, such as $P$-dependent polar Kerr and $\mu$SR experiments, are needed to ascertain whether the surface-state superconductivity in PbTaSe$_2$ has broken time-reversal symmetry \cite{47,Wahl}.

Acknowledgement: Cong Ren acknowledge stimulating discussions with Profs. Qiang-hua Wang, Hai-Hu Wen, Xin Lu, and Yan Chen. This work is supported by the National Science Foundation of China (Grant No. 11774303, 11574373,11874417,11774401) and the Strategic Priority Research Program (B) of Chinese Academy of Science (Grant No. XDB33010100). C. R. acknowledges financial support by Joint Fund of Yunnan Provincial Science and Technology Department (2019FY003008), and P.X. acknowledges financial support by National Science Foundation grant DMR-1905843.

$\ast$ cong\_ren@iphy.ac.cn

$\dag$ gfchen@iphy.ac.cn

\end{document}